%% file: main.tex
\documentclass{article} 
\usepackage{iclr2026_conference,times}
\pdfoutput=1
\input{math_commands.tex}

\usepackage{url}

\usepackage[pdftex]{graphicx} 
\usepackage{booktabs} 
\usepackage{subcaption} 

\iclrfinalcopy 
\title{TinyMusician: On-Device Music Generation with Knowledge Distillation and Mixed Precision Quantization}

\author{
Hainan Wang\textsuperscript{1,*} \quad
Mehdi Hosseinzadeh\textsuperscript{2,3} \quad
Reza Rawassizadeh\textsuperscript{1,4,*} \\
\textsuperscript{1}Metropolitan College, Department of Computer Science, Boston University \\
\textsuperscript{2}School of Engineering \& Technology, Duy Tan University, Da Nang, Vietnam \\
\textsuperscript{3}Jadara Research Center, Jadara University, Irbid, Jordan \\
\textsuperscript{4}Center of Excellence in Precision Medicine and Digital Health, Department of \\ Physiology, Chulalongkorn University, Thailand \\
\textsuperscript{*}Corresponding authors: \texttt{nanmax@bu.edu}, \texttt{rezar@bu.edu}
}

\begin{document}

\maketitle

\begin{abstract}
The success of the generative model has gained unprecedented attention in the music generation area. Transformer-based architectures have set new benchmarks for model performance. However, their practical adoption is hindered by some critical challenges: the demand for massive computational resources and inference time, due to their large number of parameters. These obstacles make them infeasible to deploy on edge devices, such as smartphones and wearables, with limited computational resources. In this work, we present \textit{TinyMusician}, a lightweight music generation model distilled from MusicGen (a State-of-the-art music generation model). TinyMusician integrates two innovations: (i) Stage-mixed Bidirectional and Skewed KL-Divergence and (ii) Adaptive Mixed-Precision Quantization. The experimental results demonstrate that TinyMusician retains 93\% of the MusicGen-Small performance with 55\% less model size. TinyMusician is the first mobile-deployable music generation model that eliminates cloud dependency while maintaining high audio fidelity and efficient resource usage. \footnote{https://github.com/maxW2000/tinyMusician}
\end{abstract}


\section{Introduction}
Music, reflecting culture, social classes, ethnic identities, and historical eras, has woven itself into humanity's shared heritage through centuries of evolution \citep{toynbee2012music}. Today, artificial intelligence (AI) has demonstrated remarkable breakthroughs across multimodal domains, including image generation \citep{liu2023aiinpaintingoverview}, inpainting, outpainting \citep{silva2024artificialaiinpainting}, short video production \citep{aiInFilm}, etc. 

Large-Language Models (LLMs) \citep{liu2023think, longTermrelaitonship1} showed excellent modeling capabilities in obtaining complex relationships in long-term contexts, which the music genre inherited. In view of this, MusicLMs \citep{agostinelli2023musiclm} and many subsequent works \citep{copet2023simpleMusicGen, lam2023efficientMusic, Suno-AI_2023} successfully applied LLMs in music generation, capitalizing on their ability to capture intricate patterns in musical sequences.

However, the pursuit of higher-quality AI music generation, driven by two technical imperatives, Scaling Law \citep{kaplan2020scalinglaw} and Emergent Capability \citep{berti2025emergentabilities}, has led to a surge in model parameters, creating critical challenges in both computation and deployment. The frequently discussed text-to-music models, for example, MusicGen-Large \citep{copet2023simpleMusicGen} and YuE-7B \citep{yuan2025yuemodel}, having undergone training on large-scale datasets, exhibited excellent capabilities in synthesizing music \citep{austin2021programlargedatasetgoodpaper}. The generated music was characterized by high fidelity \citep{yao2025jenhighfidelity}, a high level of accuracy and detail in its sound quality, and a strong coherence with the provided text prompts. Yet, this parameter escalation introduces a trilemma between musical fidelity, computational cost, and deployment feasibility, especially for on-device deployment, where there are limited computational resources, such as smartphones and extended reality glasses. The dependency on a cloud, with high computational overheads, hinders the proliferation of AI-generated products into real-world applications, such as games, and keeps these models server-dependent \citep{nocloud, huang2024largeclouddepend, ODSearch}. 

There are several efforts to reduce the model size, especially in Transformer architecture, such as efficient self-attention mechanisms \citep{attentions}. Furthermore, several approaches such as Mixture of Experts (MoE) \citep{moe}, and Low-Ranked Adaptation (LoRA) \cite{lora} are used to reduce the computational cost of feed-forward layers of transformer architecture. 

On the other hand, several approaches focus on enabling the neural network models' deployment on consumer electronics, such as Federated Learning \citep{li2020reviewfederatedlearning} and model compression techniques. Three common approaches focus on model compression and reducing the size of a neural network, including Knowledge Distillation \citep{gou2021knowledgesurvey}, pruning \citep{zhang2022advancingpruning}, and quantization \citep{wei2024quantizationreview}. 

In short, knowledge distillation transfers knowledge from a teacher model (baseline model) to a student model (smaller model), which ensures fidelity of results while having a smaller number of model parameters. This technique has demonstrated its efficacy across multiple AI domains. For instance, \citet{mullapudi2019onlinevideosdistillation} proposed JITNet, employing MRCNN \citep{tian2019mrcnn} as the teacher model, and reduced the number of parameters from 300 million to 7 million. \citet{sun2019patientBert} introduced Patient Knowledge Distillation, which distills the 12-layer BERT-original \citep{devlin2019bert} into a 6-layer BERT while preserving 97\% of its performance.

In addition to the knowledge distillation, neural network quantization has emerged as another paradigm of critical model compression \citep{gou2021knowledgesurvey, wei2024quantizationreview}. By converting high-precision model weights into compact low-bit representations without altering the network architecture \citep{gholami2022quantisurvey}, this technique has been proven effective in domains such as audio processing \citep{derrien2006newquantiinaudio}, Deep Reinforcement Learning \citep{deepRLcomp} and image processing \citep{rokh2023comprehensivequantiinimage}. 

The third common approach is pruning \citep{zhu2017prune}, which includes removing neurons that are not contributing much to the final output. Pruning could be determined based on the weight, activation value, or even the entire neuron and its connections \citep{Rezamlandai}.

While these techniques have been extensively explored in different fields, especially image recognition \citep{rokh2023comprehensivequantiinimage} and natural language processing \citep{sun2019patientBert}, their application to music generation remains underexplored. 

In this work, we present \textit{TinyMusician}, a novel lightweight model for mobile, on-device music generation, distilled from the state-of-the-art MusicGen-Small \citep{copet2023simpleMusicGen} architecture, integrating Stage-mixed Bidirectional KL-divergence in Knowledge Distillation with temperature annealing strategy \citep{zhang2024trustworthytemperatureanneling} to enhance knowledge transfer fidelity between teacher and student models. To further enhance inference efficiency, we also implement adaptive mixed-precision quantization \citep{chauhan2023postmixedprecision} and achieve 55\% reduction in model size compared to the original MusicGen with 9.5\% sacrificing melodic or harmonic fidelity. 

\begin{figure}[htbp]
    \centering
    \includegraphics[width=0.7\linewidth]{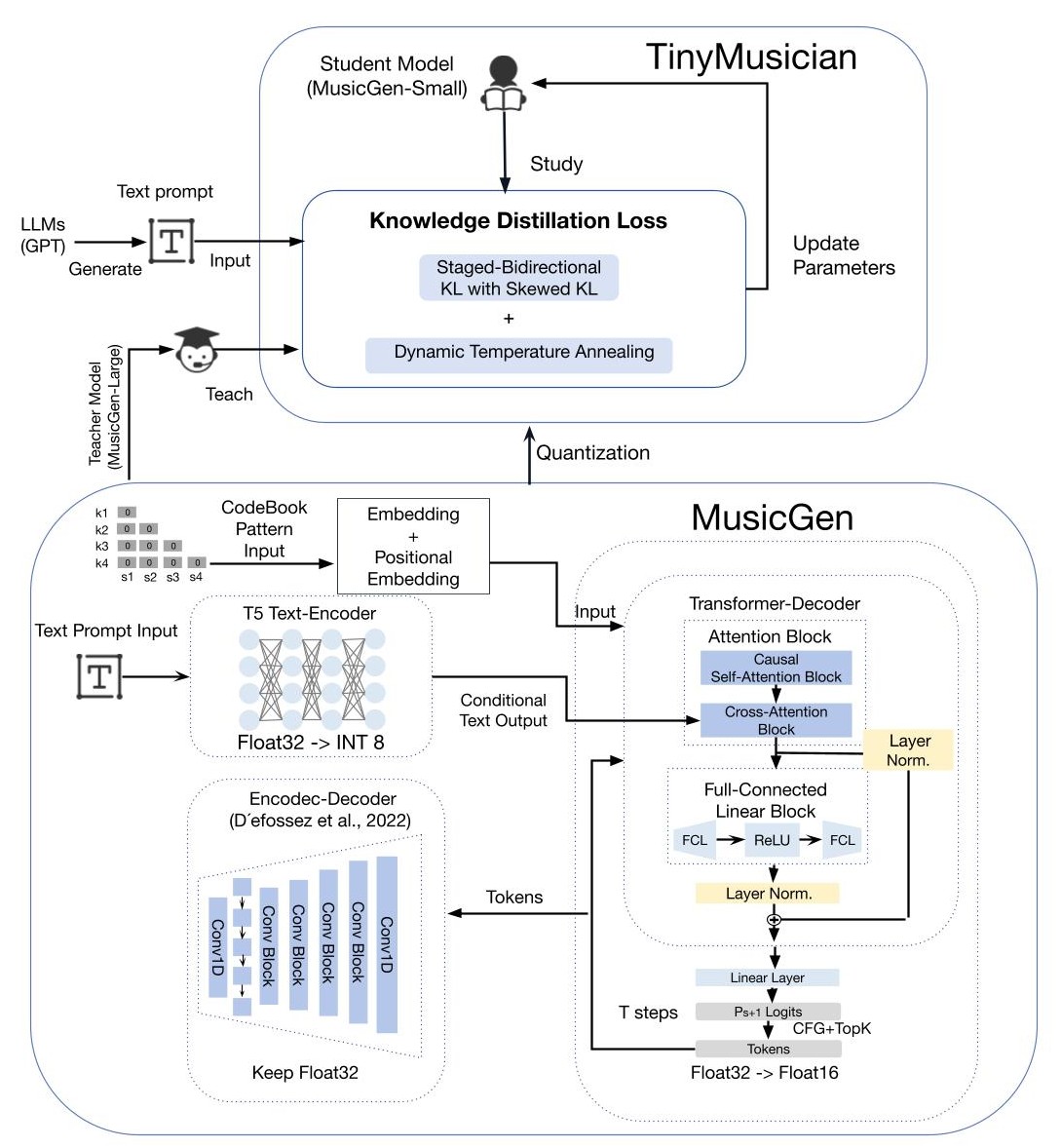}
    \caption{The architecture of TinyMusician with respect to its teacher model, i.e., MusicGen small.}
    \label{fig:tinymusicarchitecure}
\end{figure}

Figure \ref{fig:tinymusicarchitecure} presents the architecture of TinyMusician.
We have integrated TinyMusician into the iOS mobile platform through ONNX runtime conversion and platform-specific optimization for iOS and Android. To our knowledge, this is the first on-device music generation model that can run on smartphones independently of the cloud or other large computational resources.

\section{Related Works}

Deploying high-fidelity music generation on edge devices faces two co-dependent barriers: 
(1) Resource-Intensive Models: Transformer-based music models (Table \ref{tab:model_compare}) achieve remarkable quality but require strong GPU resources, which are incompatible with edge devices' constraints; 
(2) Compression Limitations: Existing compression techniques lack specialized mechanisms to preserve musical fidelity, risking perceptual degradation. 
Therefore, our related work is composed of two sections: synthetic music generation and model compression.

\subsection{Synthetic Music Generation}
Previous to neural network advances, the Markov chain \citep{hassani2016markov2, shapiro2021markov1}, rule-based models \citep{hastuti2017rulebased1, sneyers2010rulebased2}, and evolutionary algorithms \citep{loughran2020evolutionary1, kaliakatsos2020evoluation2} are three main groups of methods that are mainly used to generate music. These methods are typically parameter-based, requiring human input of parameters or configurations to guide music generation. Music generated by these methods remained quite limited.

Later, with the development of deep neural networks, generative models show incredible ability in sequential data construction, including music. Several types of generative models have been developed to meet the high-quality requirement of music generation, including RNN models \citep{goel2014polyphonicrnn1, dua2020improvedrnn2}, GAN based-models \citep{zhang2021implementgan1, huang2020emotiongan2}, and VAE based-models \citep{dhariwal2020jukeboxvae1, liang2019midisandwichvae2}. For example, Jukebox \citep{dhariwal2020jukeboxvae1}, one of the first VAE-based models, can generate full vocal music. Although the quality is limited and slow, it still demonstrates the ability to produce music that aligns with the inputs of the lyric, artist, and Genre. 

Recently, diffusion models \citep{song2020denoisingdiffusionmodelboss} and transformer-based models \citep{kang2024video2musictransformer1, shih2022themetransformer2} have emerged as the mainstream in music generation models. The learning process of diffusion models involves two core steps: a forward process that gradually adds noise to a sample and a reverse process that aims to denoise and reconstruct the original data \citep{Rezamlandai}. ERNIE-Music \citep{zhu2023erniediffusionmodel} is a diffusion-based architecture specifically designed for music, and it involves a forward step of gradually adding Gaussian noise to music waveforms and a reverse denoising process to reconstruct the original audio, using a U-Net with conditional self-attention \citep{ibtehaz2020multiresunet} to integrate text prompts from an ERNIE-M text encoder for direct text-to-waveform translation. 

Transformer-based models \citep{wen2022transformersinseriesdata}, on the contrary, are experts in modeling long-range dependencies and structural patterns, such as melodic repetition, harmonic progression, or rhythmic patterns, by processing musical sequences as tokenized events (notes, pitches, instruments) with positional encoding \citep{dash2024aitransformer-based}. However, the strong performance of transformer models also comes with a tradeoff: the transformer-based architecture requires substantial computational resources, which hinders their deployment on small battery-powered devices. 

Since our approach is also transformer-based, the popular transformer models for music generation are listed in Table \ref{tab:model_compare}. As shown in this Table, even small state-of-the-art models have a large number of parameters. For example, Yue-7B \citep{yuan2025yuemodel} has 7 billion parameters and demands about 40GB of memory model storage. Even the smallest model, MusicGen-Small, demands 10GB of GPU memory and an RTX 3080 GPU to achieve acceptable inference speeds. Large model parameters and high computational costs require devices with very high memory and computing power. 

These evidences show that directly deploying such models not only incurs significant resource costs but also triggers long inference times. Additionally, deploying these models on edge devices such as mobile phones is impossible due to their limited storage and computing resources. 

\begin{table}[]
\centering
\begin{tabular}{lccc}
\toprule
Model &  Model Size & Params & GPU (Inference)\\
\midrule
Music-LM \citep{agostinelli2023musiclm}& 3.44GB & 860M & RTX 3050 8GB \\
YuE-7B \citep{yuan2025yuemodel}& 13GB & 7B & RTX 3090 24GB \\
Flux \citep{fei2024fluxmodel} & 8.44GB & 2.1B &  RTX 3090 24GB\\ 
Musictango \citep{melechovsky2023mustango} & 5.6GB & 1.4B & RTX 3060 12GB \\ 
Spectrogram \citep{hawthorne2022multiSpectrogramDiffusion} & 1.65GB & 412M & RTX 2060 6GB \\
\bottomrule
\end{tabular}
\caption{Music generation model sizes along with GPU memory utilization}

\label{tab:model_compare}
\end{table}

\subsection{Model Compression}

A reasonable model compression method finds the balance between compressed pre-trained model memory and model performance so that the model can be deployed on various resource-constrained devices \citep{tang2024surveycompressiondefinition}.

Quantization methods have some advantages over Pruning and Knowledge Distillation. In particular, first, they are cost-effective, and most of the quantization methods don't need to retrain the entire model, making them easier for researchers with limited computing resources. Second, they support effective compression, because the weights of models from 32-bit Float to 8-bit or 4-bit Int could drastically compress model size to approximately 1/4 or 1/8. Third, quantization is highly compatible with most other model compression methods and thus flexible. 

Quantization-aware training (QAT) and Post-
Training quantization (PTQ) is a common Quantization method \citep{Rezamlandai}. QAT \citep{esser2019learnedQAT} aims to quantize the model during the training phase, while PTQ \citep{shang2023postPTQ} considers the quantization after training. Due to the cost-effectiveness of time and computational resources, PTQ is more popular.  Most PTQ approaches quantize parameters in weights and activations in each layer, and can be divided into three subsets: Weight-only Quantization, Key-Value (KV) Cache Quantization, and Weight-activation Quantization \citep{liu2025quantizationsubsets}. PTQ advancements have reshaped model efficiency. For example, GPTQ \citep{frantar2022gptq} is a Weight-only method that can compress popular open-source models down to 3 and 4 bits. SmoothQuant \citep{xiao2023smoothquant}, in contrast, introduces joint weight-activation quantization, balancing their dynamic ranges to reduce error propagation in vision transformers. While PTQ methods have improved model efficiency, their application to music generation models remains underexplored. 

Unlike text or images, music synthesis demands precise preservation of temporal dynamics and spectral fidelity, which are highly sensitive to quantization errors in both weights and activations. Applying uniform quantization across all model weights and activations, as commonly done in other domains, risks significant degradation in musical quality \citep{lohar2023soundmixedprecision}. To address these challenges, for music generation, a mixed-precision quantization approach is essential. In contrast, \citet{xiao2023smoothquant} introduce joint weight-activation quantization, balancing their dynamic ranges to reduce error propagation in vision transformers. While PTQ methods have improved model efficiency, their application to music generation models remains underexplored. Unlike text or images, music synthesis demands precise preservation of temporal dynamics and spectral fidelity, which are highly sensitive to quantization errors in both weights and activations. 

Applying uniform quantization across all model weights and activations, as commonly done in other domains, risks significant degradation in musical quality. To address these challenges, a mixed-precision quantization approach is essential, and our model, \emph{TinyMusician} implements it.

Knowledge Distillation (KD), based on how to design the loss function, can be further categorized into two groups: logit-based KD, and feature-based KD \citep{hinton2015distillingfirst}. Logit-based KD typically uses KL-Divergence or Mean Square Error (MSE) to minimize the logits between the teacher and the student. DistilBERT \citep{sanh2019distilbert}, for example, is a KD of BERT, which is 40\% smaller, 60\% faster, retains 97\% of BERT's language understanding capabilities, and is trained with a triple loss during pre-training, demonstrating its effectiveness in various downstream tasks. \citet{schmid2023audioefficient} propose an offline KD training method from high-performance yet complex Transformer models to efficient CNN models. It constructs different audio tagging models with different complexities, outperforming previous solutions in terms of model size, computational efficiency, and prediction performance, assessed via Frechet Audio Distance (FAD) \citep{kilgour2018frFAD}, which quantifies audio by comparing feature distributions, and CLAP scores \citep{ye2023clapspeech}, which measure text-audio semantic alignment via contrastive learning, and achieving a new single-model state-of-the-art mean average precision of 0.483 on the AudioSet dataset.

Feature-based KD aims to minimize the intermediate features between the teacher and the student. PKD \citep{sun2019patientfeaturebased} introduced MSE as a loss function and proposed two strategies: the student learns the last few layers in the teacher, and the others learn every two layers' representations of the teacher. MT-BERT \citep{wu2021onemultiteacher} method uses multiple teacher pre-trained language models with a new finetuning framework and new loss functions to better compress PLMs and outperforms single-teacher and some multi-teacher distillation methods. However, as with Quantization, KD is also still underexplored in the music generation area.

\section{TinyMusician}

As it has been stated before, in addition to knowledge distillation, TinyMusician introduces two salient novelties to enable on-device music deployment, which we describe in this section. 

\subsection{Knowledge Distillation with Stage-mixed Bi-directional and Skewed KL} 
To perform knowledge distillation, we choose MusicGen-Large as the teacher model \citep{copet2023simpleMusicGen}, and apply our knowledge distillation on MusicGen-Small, as the student model, and further improve it, which leads to the TinyMusician. Traditional one-directional KL-Divergence aims to force the student model to mimic the output distribution of the teacher model. However, music should keep chronological coherence and local tone detail; thus, inspired by the methodology proposed by \citet{yang5219975modalklsourcestaged}, we introduced an improved formulation of Bidirectional KL-Divergence, called \textit{Stage-mixed Bidirectional KL-Divergence}, as a loss function and conducted comparative experiments against traditional KL Divergence variants. The detailed analysis of different divergence metrics and experimental configurations will be presented in Section \ref{sec:experiments}. The definition of Stage-mixed Bidirectional and Skewed KL-Divergence is presented in Equation \ref{eq:stage_kl_loss}. 
\begin{equation}
\begin{split}
\mathcal{L}_{\text{KL}}(t) &= \alpha(t) \cdot \left[ \gamma_1 D_{\text{KL}}(T \| S) + (1 - \gamma_1) D_{\text{KL}}(T \| S_{\lambda}) \right] \\
&\quad + (1 - \alpha(t)) \cdot \left[ \gamma_2 D_{\text{KL}}(S \| T) + (1 - \gamma_2) D_{\text{KL}}(S \| T_{\lambda}) \right]
\end{split}
\label{eq:stage_kl_loss}
\end{equation}
where the mixed distributions are defined as:
\begin{align}
S_{\lambda} &= \lambda T + (1 - \lambda)S \\
T_{\lambda} &= (1 - \lambda)T + \lambda S
\end{align}
In Equation \ref{eq:stage_kl_loss}, $\mathcal{L}_{\text{KL}}(t)$ represents the stage-mixed KL-Divergence loss at time step $t$. $\alpha(t)$ is a dynamic weight function that varies with the time step $t$, which is used to adjust the proportion of different KL-Divergence terms in different stages. $\gamma_1$ and $\gamma_2$ are hyperparameters. $\gamma_1$ is used to balance the two KL-Divergence terms in the first part of the equation, and $\gamma_2$ is used for the second part. $T$ represents the teacher distribution, and $S$ represents the source distribution. $\mathcal{D}_{\text{KL}}(A || B)$ represents the KL-Divergence between distribution $A$ and distribution $B$. The mixed distributions $S_{\lambda} = \lambda T + (1-\lambda)S$ and $T_{\lambda} = (1-\lambda)T + \lambda S$ represent convex combinations of the teacher and student distributions, where $S_{\lambda}$ smooths the student's output for stable forward KL Divergence optimization, while $T_{\lambda}$ robustifies the teacher's reference for resilient reverse KL Divergence learning.

\begin{equation}
\alpha(t) = 
\begin{cases} 
1 & \text{if } t < \tau_{\text{step}} \\
0 & \text{otherwise}
\end{cases}
\label{eq:dynamic_weight_new}
\end{equation}
In Equation \ref{eq:dynamic_weight_new}, $\alpha(t)$ is a dynamic weight function that depends on the current time step $t$ and a pre-defined step threshold $\tau_{\text{step}}$. When the time step $t$ is less than the threshold $\tau_{\text{step}}$, $\alpha(t)$ takes the value of 1, the first part of the loss equation, i.e., \(\left[\gamma_1 \mathcal{D}_{\text{KL}}(T || S)+(1 - \gamma_1)\mathcal{D}_{\text{KL}}(T || S_{\lambda})\right]\) is taken into the account, while the second part, i.e., \(\left[\gamma_2 \mathcal{D}_{\text{KL}}(S || T)+(1 - \gamma_2)\mathcal{D}_{\text{KL}}(S_{\lambda} || T)\right]\) weights 0 and is thus ignored. When \(t\geq\tau_{\text{step}}\), \(\alpha(t)\) is 0. In this case, the first part of the loss equation is ignored, and the second part is fully taken into account.

To meet the requirement of KD in different stages, we design an \textit {adaptive temperature annealing mechanism} inspired by the strategy proposed by \citet{manvi2024adaptiveannelingtemprature}. Unlike the exponential annealing schedule proposed in their work, our approach employs a linear decay. This adaptive temperature annealing is straightforward and scalable because it avoids the complex parameter tuning required by nonlinear schedules, while still effectively balancing exploration and exploitation in the generation process. Equation \ref{eq:temperature_formula} formalizes our adaptive temperature annealing approach.
\begin{equation}
\tau = T_b - (T_b - T_f) \times \left(\frac{s}{L_{max}}\right)
\label{eq:temperature_formula}
\end{equation}
Here, $T_b$ represents the initial temperature, $T_f$ represents the final temperature, $s$ represents the current step, and $L_{max}$ represents the maximum output length.

\subsection{Customized Quantization} In addition to our proposed KD, we adopt a post-training \citep{shang2023postPTQ} mixed-precision method to quantize the MusicGen-small model. The MusicGen model can be partitioned into three distinct components: the T5 Text-Encoder \citep{ni2021sentencet5},  the MusicGen-Decoder, and the Encodec-Decoder \citep{defossez2022highencodecdecoder}. Each of these components is quantized into different formats: specifically, the Text-Encoder is quantized to Int8 to balance efficiency and representation preservation; the MusicGen-Decoder is quantized to Float16 to maintain autoregressive generation stability; and the Encodec-Decoder is kept in Float32 to ensure high-fidelity audio reconstruction. 

The Text Encoder takes text as input embeddings, produced by a tokenizer or embedding layer, and outputs the last hidden states. In MusicGen-Decoder, the transformer performs autoregressive token generation by processing a sequence of discrete tokens, step-by-step. At each step, it uses causal self-attention to focus only on previously generated tokens, ensuring no future information is accessed. It also incorporates conditional signals (the text embeddings) via cross-attention to guide the generation. Based on these inputs, the transformer predicts the next token (using Classifier Free Guidance (CFG) \citep{sanchez2023stayCFG} strategy and the top-k sampling strategy to guide the model's output) in the sequence, which is then added to the existing sequence. This iterative process continues until a complete token sequence is generated, and each new token builds on the context of all prior ones. 

Lastly, the generated tokens are then fed into a subsequent Encodec-Decoder module. The decoder within this module further decodes these intermediate tokens into raw audio waveforms, completing the end-to-end text-to-music generation pipeline. Quantization efficacy and latency/quality trade-offs are evaluated in Section \ref{sec:results}, demonstrating minimal degradation compared to full-precision baselines.

\section{Experiments}
\label{sec:experiments}
\subsection{Dataset}
We conduct our experiments on the MusicCap Dataset \citep{lee2023annotatormusiccap}, which is a large-scale dataset for music-text alignment tasks. It consists of 5,500 high-quality music-text pairs. Each sample is annotated with two types of descriptions: (i) a list of English-language musical aspects, which captures elements such as genre, tempo, and instrumentation; and (ii) a free-form text caption authored by professional musicians, offering qualitative insights into the musical content. 

\subsection{Experimental Setup}
All experiments to train the model are performed on a hardware environment equipped with an RTX 4090 GPU (24GB), a 16 vCPU Intel(R) Xeon(R) Gold 6430 CPU, Pytorch version 2.5.1, and the operating system is Ubuntu 18.04.

\subsection{Model Training}
For knowledge distillation, we used GPT-4o to generate 200 music-related texts as prompts that guide the logits generation of both student and teacher models, and we separate the datasets into train, validation, and test. 

These prompts are meticulously crafted and include multi-dimensional musical attributes, such as temporal diversity, Genre, and instrumentation, and emotional and semantic nuances. MusicGen-small (the student model) was trained on GPU for approximately 30 hours for 10 epochs. To better show the performance of our loss function, we also trained the model when the loss function was set as Stage-mixed Bidirectional and Skewed KL Divergence, and other KL Divergence methods (as shown in Table \ref{tab:kl_methods}), and we compare the loss score of these functions.

\begin{table}[]
\centering
\footnotesize
\setlength{\tabcolsep}{3pt}
\renewcommand{\arraystretch}{0.8}
\begin{tabular}{cccc}
\toprule
\textbf{Method} & \textbf{Type} & $\lambda$ \textbf{Sch.} & \textbf{Obj. Func.}  \\
\midrule
Forward KL \citep{jerfel2021variationalfwkl} & Forward & -- & $\mathbb{E}_P[\log(P/Q)]$  \\ 
Backward KL \citep{malinin2019reversebackwardKL} & Backward & -- & $\mathbb{E}_Q[\log(Q/P)]$  \\
Fixed-Param BiKL \citep{bai2024blinfixedparakldi}  & Bidirectional & Constant & $\lambda_{\text{fix}}$(KL\textsubscript{F} + KL\textsubscript{R}) \\
BiKL \citep{li2024bildnormalkl} & Bidirectional & $\lambda$=1 & KL\textsubscript{F} + KL\textsubscript{R} \\
Stepped BiKL \citep{yang5219975modalklsourcestaged} & Bidirectional & Adaptive & $\lambda(t)$KL\textsubscript{F} + [1-$\lambda(t)$]KL\textsubscript{R} \\
\bottomrule
\end{tabular}
\caption{KL Divergence Method. a) Forward KL: Forward KL Divergence, b) Backward KL: Backward KL Divergence, Fixed-Param BiKL: Fixed-Parameter Bi-directional KL Divergence, BiKL: Bi-directional KL Divergence, and Stepped BiKL: Stepped Bi-directional KL Divergence. Specifically, Forward and Backward KL Divergence only have one direction. Bi-directional KL has both forward and backward, yet it realizes this through three distinct methods, as presented in the table (where the formulas in the “Obj. Func.” column correspond to these different realization logics)}
\label{tab:kl_methods}
\end{table}

To perform the end-to-end conversion from PyTorch to ONNX format, after the knowledge distillation, we use the Optimum-Cli tool \footnote{https://github.com/huggingface/optimum}, for model optimization and deployment. This tool is chosen due to its native support for mixed-precision workflows and automated graph optimization, ensuring compatibility with downstream inference engines. We compare the performance of the various formats, include the original Torch version and the ONNX version with different quantization. 

\subsection{Mobile Device Deployment} After our model has been built, to deploy it on edge devices, we convert the model format from PyTorch to Open Neural Network Exchange (ONNX)\footnote{https://github.com/onnx/models}. ONNX is an open-source format for neural network models \citep{shridhar2020interoperatingonnx}.
To evaluate the model's performance in real-world scenarios, we deploy the converted ONNX model to on-device testing environments. The device we chose is the iPhone 16 Pro, operating on iOS 18.2, which is equipped with an A18 Pro processor with a 6-core GPU, and 8GB of RAM. We use ONNXRuntime to execute the ONNX model on edge devices. The running screenshot and codes are displayed in the appendix \ref{appendix:ondevice}.

\subsection{Ablation Study}
To investigate the individual and combined effects of knowledge distillation and quantization, we design four configurations based on MusicGen-Small, all evaluated on the MusicCap dataset \citep{lee2023annotatormusiccap} with consistent metrics as specified in subsection:

\textit{Baseline}: Original MusicGen-Small model without knowledge distillation or quantization, maintaining the default architecture and parameters of the base model.

\textit{MusicGen Small with KD}: MusicGen-Small integrated with Stage-mixed Bidirectional KL-Divergence distillation (using MusicGen-Large as the teacher model) but without quantization, focusing on the impact of knowledge transfer alone.

\textit{MusicGen Small (Quantization)}: Original MusicGen-Small applied with adaptive mixed-precision quantization (targeting different components like Text-Encoder and MusicGen-Decoder) but without knowledge distillation, isolating the effect of quantization on efficiency and quality.

\textit{TinyMusician (KD + Quantization)}: Full TinyMusician framework, combining both Stage-mixed Bidirectional KL-Divergence distillation and adaptive mixed-precision quantization to evaluate the synergistic effect of the two techniques.

\subsection{Comparison with State-of-the-Arts}

We evaluate TinyMusician's performance across different formats and state-of-the-art AI music generation models, including YuE \citep{yuan2025yuemodel}, DiffRhythm \citep{ning2025diffrhythm}, InspireMusic \citep{InspireMusic2025}, CRFM \citep{thickstun2023anticipatorycrfm}, Magenta-Realtime \citep{magenta_rt}, Musicgen-Small \citep{copet2023simpleMusicGen}. 

In particular, we incorporate resource utilization metrics and accuracy-related established benchmarks. The resource utilization includes inference time (in seconds), GPU FLOPS, CPU utilization (in percentage), Memory Usage (in GB), GPU Memory Usage (in GB), and model size (in gigabytes), which collectively characterize the models' computational efficiency and resource requirements. 

To measure accuracy-related benchmarks, we employ two important benchmarks, i.e., CLAP (Contrastive Language Audio Pretraining) \citep{ye2023clapspeech} and FADscore (Frechet Audio Distance Score) \citep{kilgour2018frFAD}. CLAP is a framework that can capture the semantic relationship between audio and text. FADscore, on the other hand, measures the similarity between the distribution of generated audio and the real-world audio distribution. 

\section{Results and Discussion}
\label{sec:results}
In this section, we present and analyze the experimental results to validate the effectiveness of TinyMusician.

We first examine the training dynamics of different loss functions to highlight the advantages of our proposed \textit{Stage-mixed Bidirectional KL divergence}. Next, we report the findings of the ablation study, which quantifies the individual and combined impacts of knowledge distillation and quantization on model performance and efficiency. Finally, we conduct a detailed comparison with state-of-the-art music generation systems, demonstrating the superior trade-off between efficiency and generation quality achieved by TinyMusician.

\subsection{Training Dynamics of Loss Functions}

\begin{figure}
    \centering
    \includegraphics[scale=0.4]{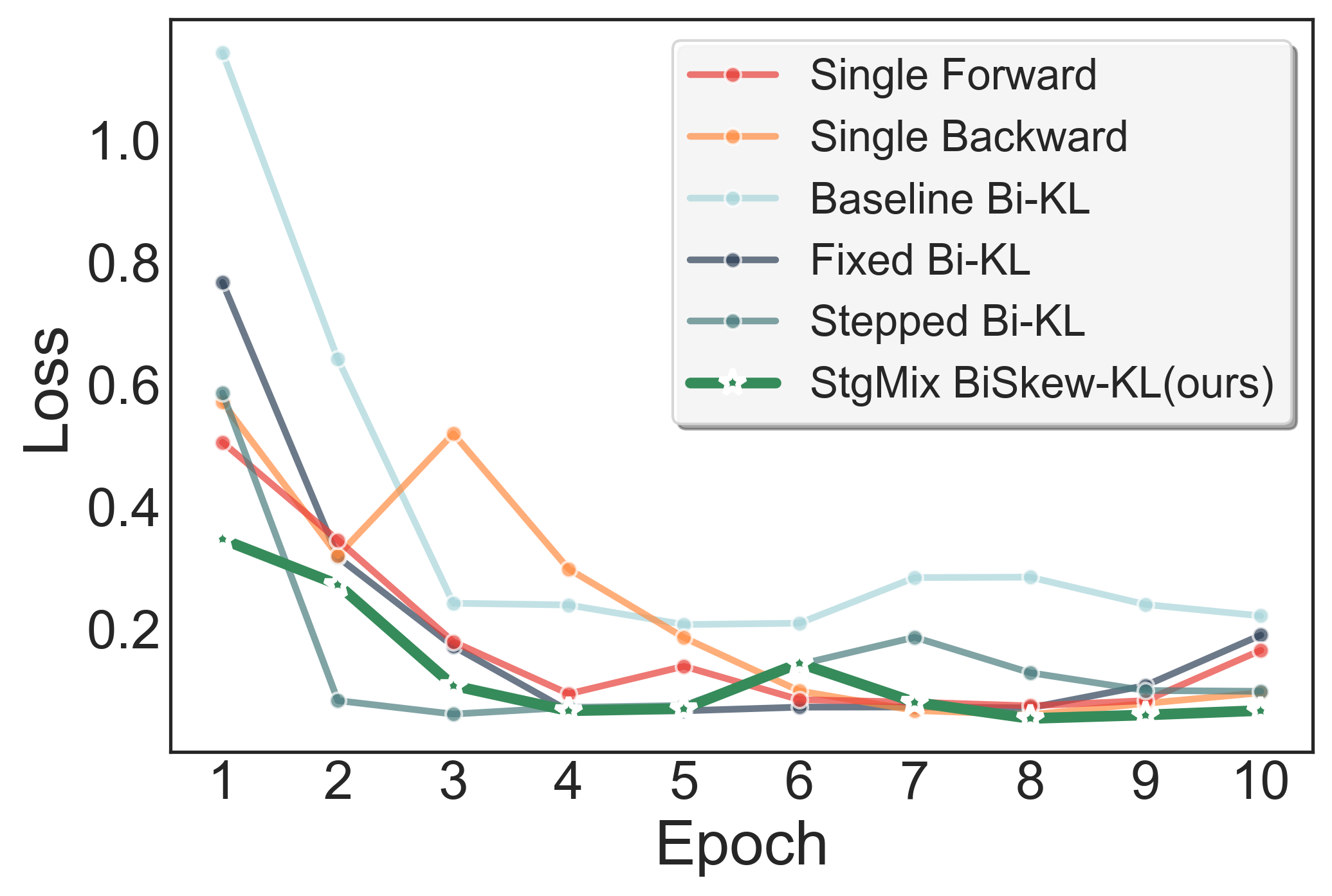} 
    \caption{Comparison of Training Loss on different KL-Divergence Methods} 
    \label{fig:kl_loss_comparison_train} 
\end{figure}

The results presented in Figure \ref{fig:kl_loss_comparison_train} show that our proposed  \textit{Stage-mixed Bidirectional KL divergence} loss function stabilizes training dynamics and enhances model generalization.

The superior performance of our method is characterized by smooth training loss descent, minimal validation loss fluctuations, and the lowest final loss ($\approx 0.13$). Unlike single-directional KL divergence (Forward/Backward), which suffers from late-stage instability, our bidirectional formulation balances these two directions through dynamic weighting ($\alpha(t)$).

Furthermore, our method avoids the dramatic oscillations of \textbf{Fixed bi-KL divergence} and \textbf{Baseline Bi-KL divergence}, highlighting its strong generalization ability. 

\begin{figure}[htbp]
    \centering
    \includegraphics[scale=0.35]{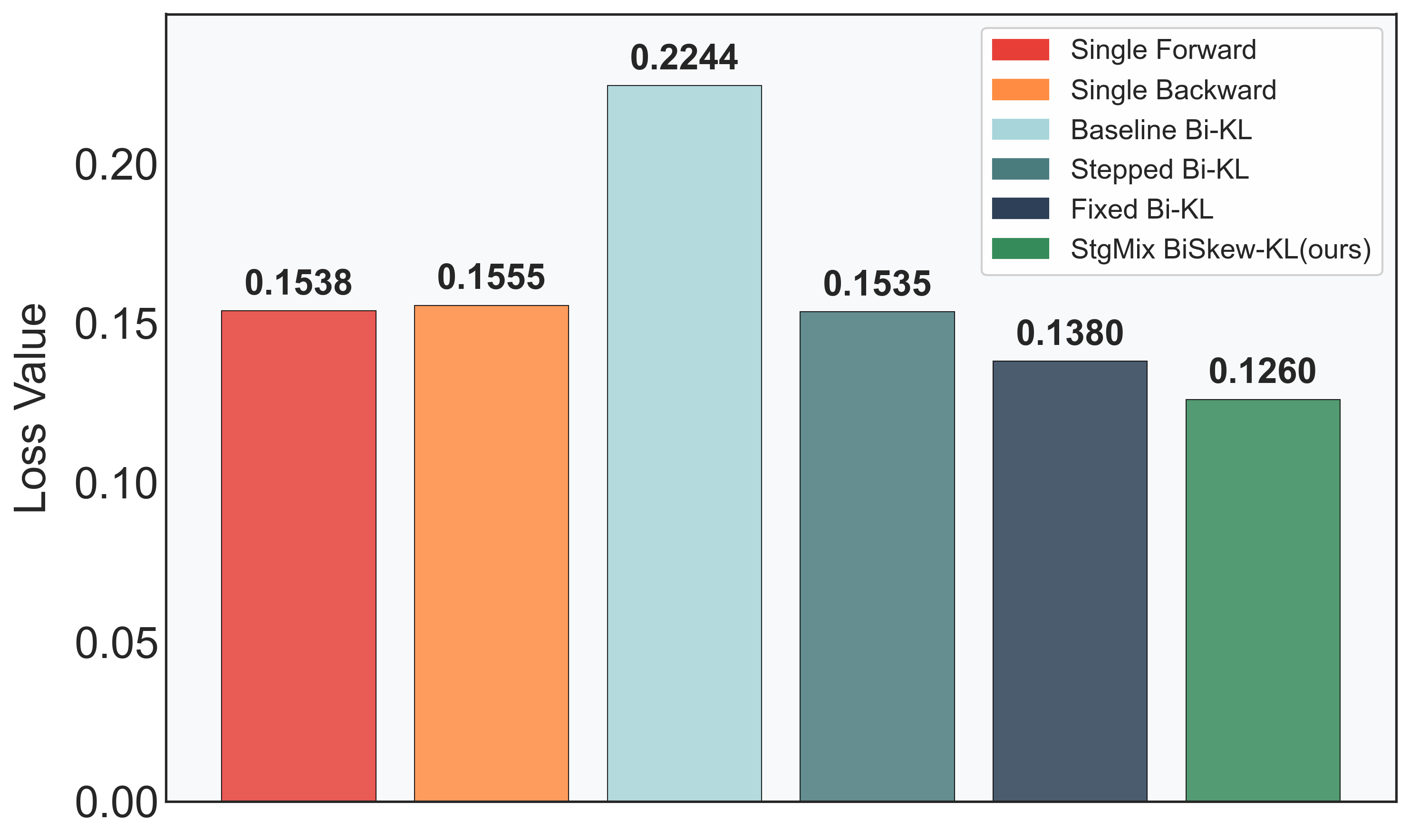} 
    \caption{Comparison of Test Loss} 
    \label{fig:kl_loss_comparison_test} 
\end{figure}

This is particularly valuable for music generation tasks, as overfitting to training data (indicated by volatile validation loss) would lead to inconsistent audio quality, such as abrupt shifts in melody or rhythm. Lastly, even though Stepped Bi-KL shows a similar pattern and performance, our method still demonstrates better results in the late stages. The results shown in Figure \ref{fig:kl_loss_comparison_test} (b) present the superiority of our method. Among all compared KL Divergence formulations, our bidirectional KL Divergence with dynamic weighting achieves the lowest test loss. This not only indicates more stable training convergence but also reflects stronger generalization capability. 

This performance originates from two synergistic mechanisms:  
(i) In the early training stages, the model prioritizes learning the teacher's overall structural patterns by emphasizing the divergence from the teacher to the student's smoothed output. As training progresses beyond a predefined threshold ($\tau_{\text{step}}$), the focus shifts to refining local temporal details critical for music—such as rhythmic consistency and melodic flow—by emphasizing the divergence from the student to the teacher's adjusted output. This stage-specific adaptation of weight distribution ($\alpha(t)$) is governed by a dynamic coefficient that transitions from 1 to 0 at $\tau_{\text{step}}$, ensuring the model balances global pattern learning and local detail preservation during different training phases. 

(ii) The use of blended distributions (combining teacher and student outputs) prevents the student from overfitting to the teacher's specific non-generalizable patterns, while creating a stable reference frame for capturing long-range musical dependencies like harmonic progressions or thematic repetitions.

\subsection{Ablation Study}
To evaluate how Knowledge Distillation (KD) and Quantization shape the performance and efficiency of TinyMusician among MusicGen-Small, we conduct an ablation study by isolating their individual impacts and analyzing their combined effect.

\subsubsection{Impact of Knowledge Distillation}
Table \ref{tab:model_ablation_scores} shows the result of accuracy. KD marginally improves generation quality (FAD score drops from 6.49 to 6.44) but slightly degrades text-audio alignment (CLAP score falls from 0.303 to 0.301). This suggests KD preserves fine-grained audio details from the teacher model but may dilute text-guided conditioning. Notably, KD alone does not change the model’s efficiency in traditional metrics: our measurements (consistent with Figure \ref{fig:ablationstudy_music_comparison}) show that compared to the baseline MusicGen-Small, the TinyMusician with KD-optimized retains nearly identical inference latency and memory footprint. This stability in efficiency metrics is likely due to the preservation of the baseline’s architectural backbone during distillation, where knowledge transfer focuses on refining output quality rather than reducing model size or computational complexity.

\begin{table}
\centering
\small
\begin{tabular}{@{}lcc@{}}
    
    \toprule
    \textbf{Model}  & \textbf{FADscore} $\downarrow$ & \textbf{CLAPscore} $\uparrow$  \\
    \midrule
    MusicGen-Small (Baseline) & 6.49 & 0.303 \\
    TinyMusician & \textbf{6.44} & 0.301 \\
    MusicGen-Small + Quantization & 7.11 & \textbf{0.352} \\
    TinyMusician + Quantization & 7.05 & 0.343 \\
    \bottomrule
    
\end{tabular}
\caption{The Scores of Ablation Study. $\boldsymbol{\text{FADscore} \downarrow}$ represents lower is better, $\boldsymbol{\text{CLAPscore} \uparrow}$ represents higher is better}
\label{tab:model_ablation_scores} 
\end{table}

\subsubsection{Impact of Quantization}
\begin{figure}
    \centering 
    \includegraphics[scale=0.35]{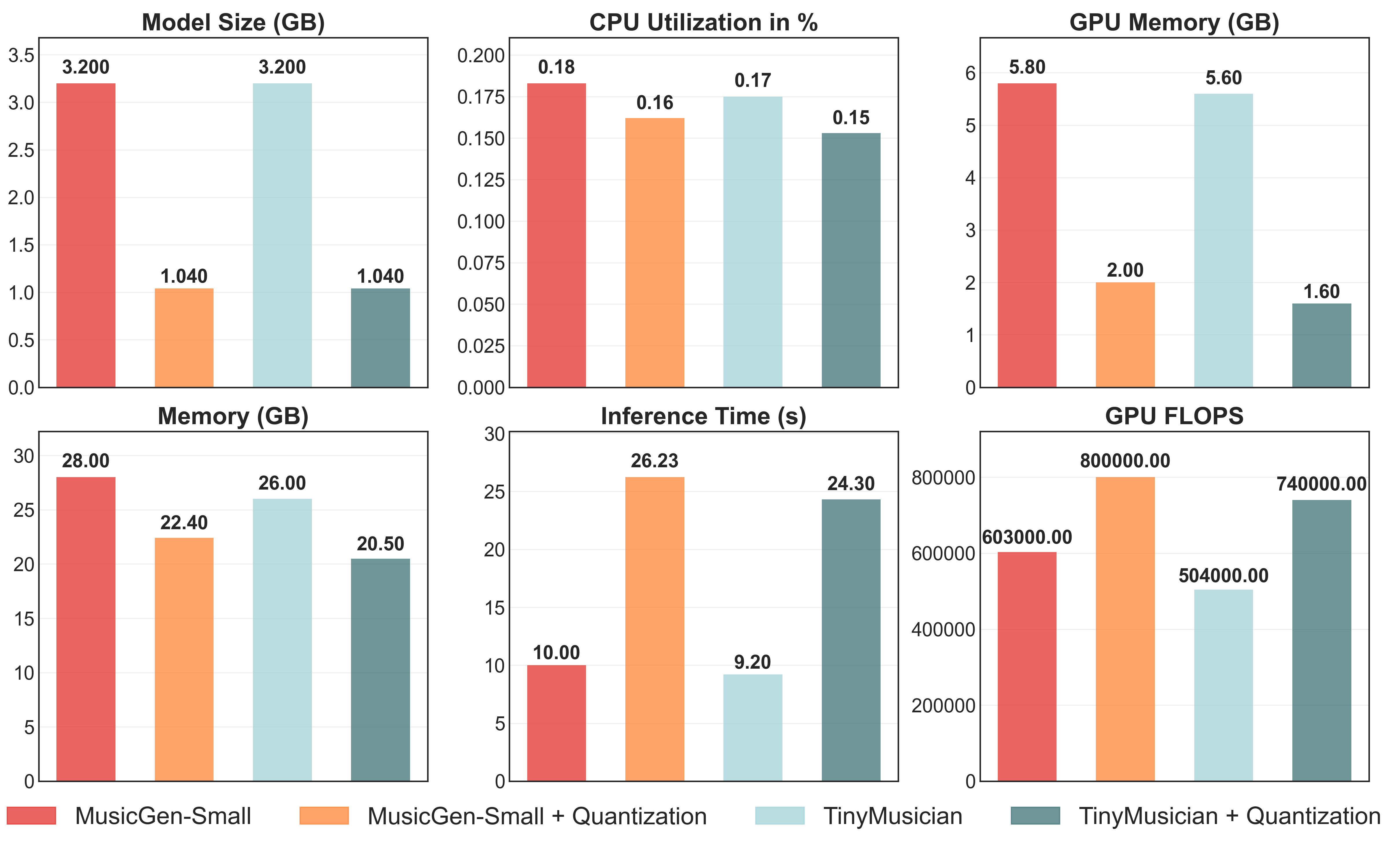} 
    \caption{Resource Utilization Comparison between TinyMusician and MusicGen-Small with different configurations.} 
    \label{fig:ablationstudy_music_comparison} 
\end{figure}

As shown in Table \ref{tab:model_ablation_scores}, quantization drastically boosts text-audio alignment (CLAP score jumps to 0.352) but harms generation quality (FAD score rises to 7.11). This trade-off is initially counterintuitive, as quantization typically reduces model precision. Quantization acts as a form of regularization \citep{moradi2020surveyregularization}, constraining model complexity and reducing overfitting \citep{ying2019overviewoverfitting} on the alignment task. When combined, KD and quantization strike a balance: the joint approach achieves a FAD score of 7.05 (better than quantization alone) and retains strong alignment (CLAP score 0.343). As shown in Figure \ref{fig:ablationstudy_music_comparison}, quantization delivers extreme compression (model size shrinks from 3.2GB to 1.04GB, GPU memory from 5.8GB to 2GB), while KD adds negligible overhead. However, the joint method inherits quantization’s latency penalty (26.54s vs. 10s for the baseline), likely due to unoptimized quantized kernels on our test hardware. These results underscore the potential of hybrid optimization strategies, contingent on hardware-aware deployment.

\subsection{Comparison with State-of-the-Art Models}

\begin{figure}[htbp]
    \centering 
    \includegraphics[scale=0.25]{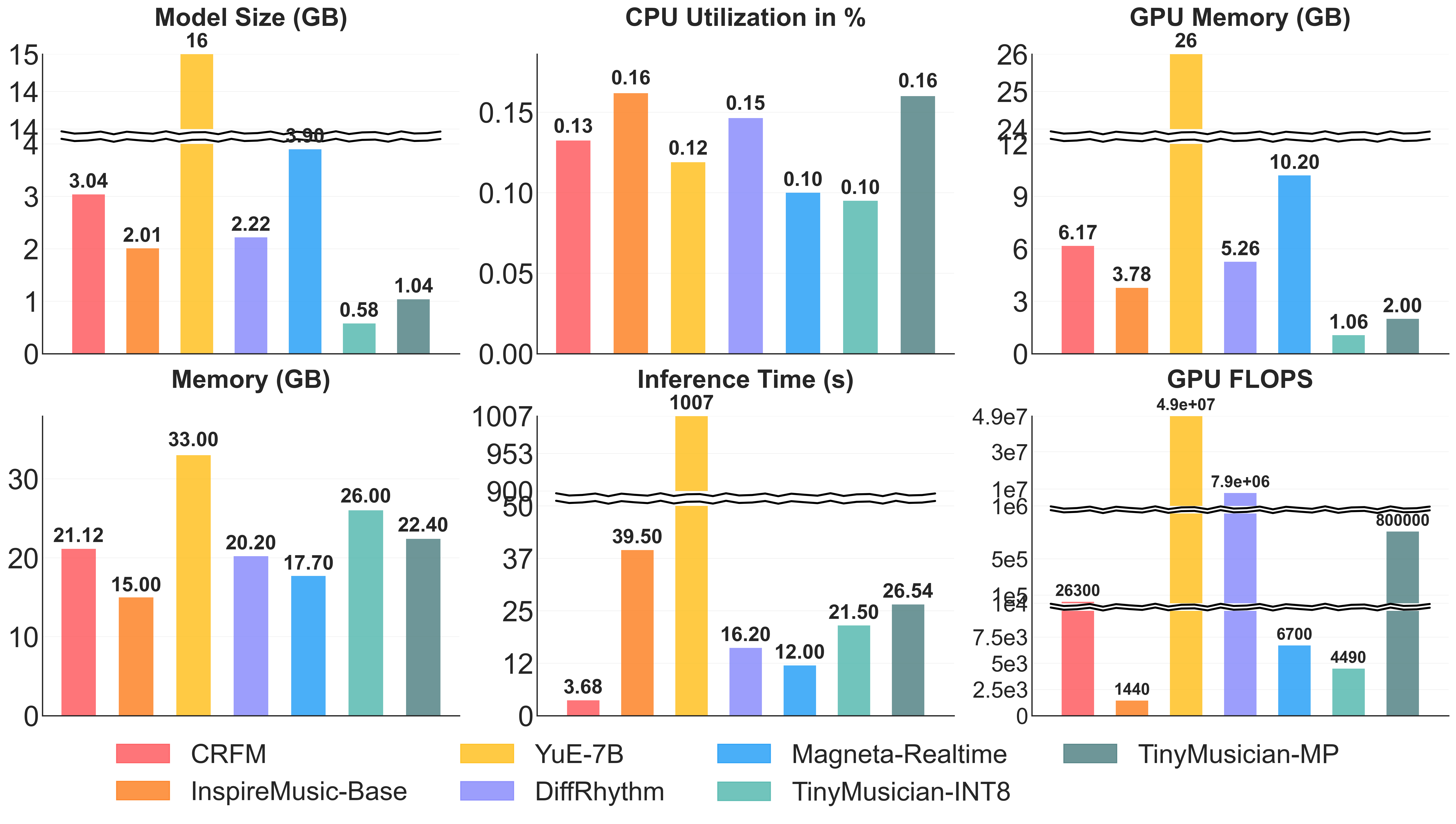} 
    \caption{Resource utilization comparison among different models. TinyMusician-MP: TinyMusician-MixedPrecision.}
    \label{fig:model_format_consumption} 
\end{figure}

Table \ref{tab:model_format_acc} and Figure \ref{fig:model_format_consumption} reflect the performance of models across various dimensions, including model size, resource consumption (CPU/GPU utilization, memory), inference efficiency (time, FLOPs), and quality metrics (FADscore \citep{kilgour2018frFAD}, CLAPscore \citep{ye2023clapspeech}). Focusing on the state-of-the-art models, MusicGen-Small quantization variants, the MusicGen-Small/ONNX(KD) Mixed configuration demonstrates outstanding trade-off advantages.  

Large-scale models (e.g., YuE-7B, DiffRhythm) prioritize fidelity but suffer from prohibitive size and latency, while pure Int-8 quantization (0.58GB) sacrifices musical coherence (e.g., abrupt rhythm shifts). In contrast, the TinyMusician-MixedPrecision approach strikes a critical balance: its 1.04GB footprint retains near-baseline fidelity with FADscore of 7.05, approaching full-precision’s 6.44, and outperforms all competitors with CLAPscore of 0.373, underscoring stronger text-music alignment. Though its 26.54s latency marginally exceeds INT8, it remains orders of magnitude faster than bulky models (e.g., YuE-7B’s 1007s). The MusicGen-Tiny variant further validates scalability: with 40\% fewer parameters, it nears baseline fidelity, demonstrating the framework’s potential for efficiency-driven refinement.

This hierarchy highlights a core insight: naive compression (Int-8) or scale (large models) fails to serve on-device music generation, whereas our TinyMusician-MixedPrecision strategy harmonizes compactness with perceptual quality — a critical requirement for edge deployment.

\begin{table}
\centering
\small
\begin{tabular}{@{}lcc@{}}
    
    \toprule
    \textbf{Model} & \textbf{FADscore} $\downarrow$ & \textbf{CLAPscore} $\uparrow$  \\
    \midrule
    CRFM  & 8.9 & 0.222 \\
    InspireMusic-Base & 10.7 & 0.150 \\
    YuE-7B  & 10.41 & 0.310 \\
    DiffRhythm  & 10.97 & 0.16 \\
    Mageneta-Realtime  & \textbf{6.79} & 0.311\\
    MusicGen-Small (Baseline) & 6.49 & 0.303 \\
    TinyMusician & 6.44 & 0.301 \\
    TinyMusician-Int8  & 8.30 & 0.283\\
    TinyMusician-MixedPrecision & 7.05 & \textbf{0.373} \\
    \bottomrule
    
\end{tabular}

\caption{TinyMusician model performance in comparison to state-of-the-art models.}
\label{tab:model_format_acc} 
\end{table}

\section{Conclusion}
In this study, we address the critical challenge of deploying large music generation models on resource-constrained edge devices, such as mobile phones, by introducing TinyMusician, a lightweight framework that integrates knowledge distillation and adaptive mixed-precision quantization. By using MusicGen as a baseline, we propose a stage-mixed bidirectional KL Divergence loss with a dynamic temperature annealing strategy to enhance the performance of knowledge transfer between the teacher and student models. To further optimize inference efficiency on devices, we apply mixed-precision quantization to different components of the MusicGen model, achieving a 55\% reduction in model size while preserving the performance. 
For future work, we plan to investigate how to speed up on-device inference by using the different model formats. Also, could apply this framework to the state-of-the-art generative models and conduct more experiments to optimize further compression strategies while saving the output quality.

\bibliography{ref}
\bibliographystyle{iclr2026_conference}

\newpage
\appendix
\section{Appendix}
\label{appendix:ondevice}
\begin{figure}[htbp]
    \centering
    \begin{subfigure}[b]{0.3\textwidth} 
        \centering
        \includegraphics[width=\textwidth]{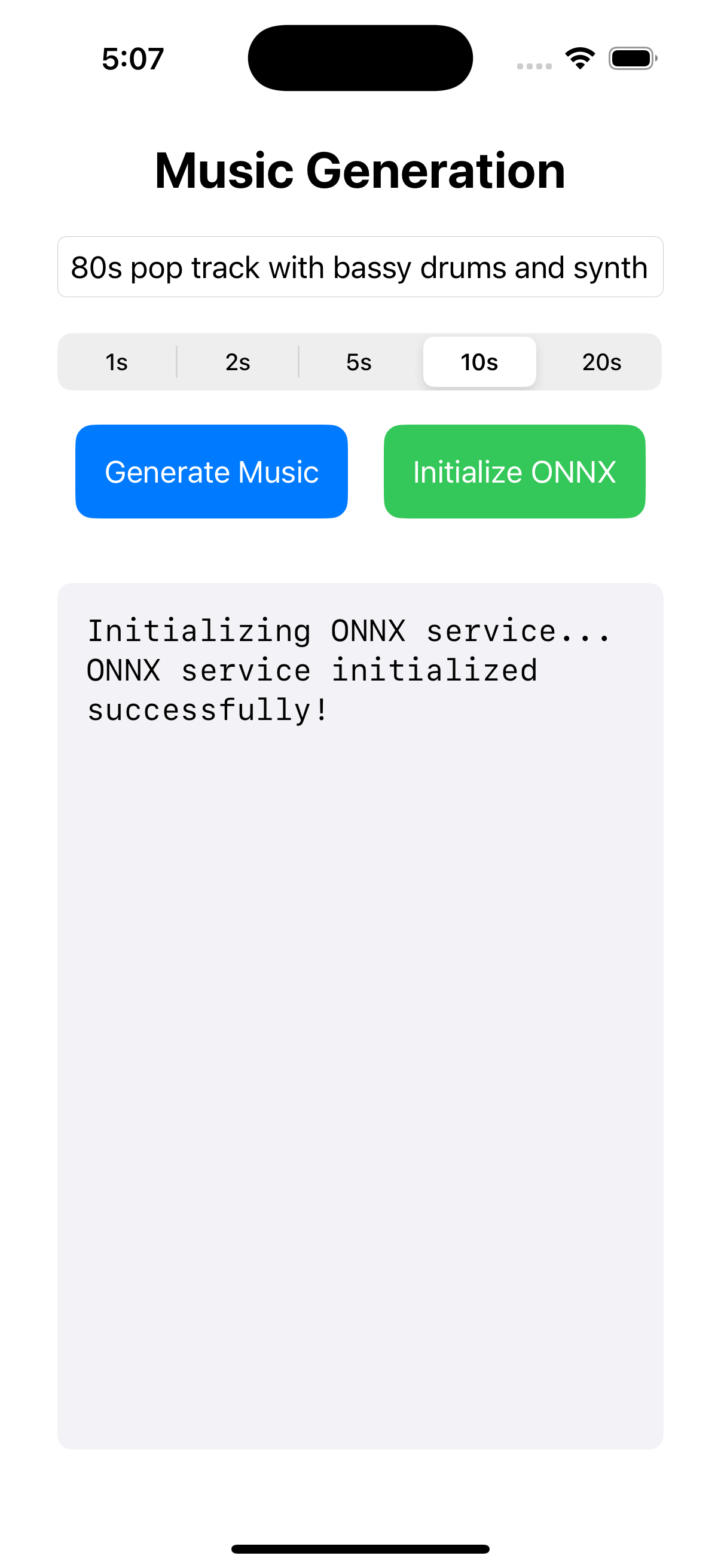} 
        \caption{Main Screen} 
        \label{fig:sub1}
    \end{subfigure}
    \hfill
    \begin{subfigure}[b]{0.3\textwidth}
        \centering
        \includegraphics[width=\textwidth]{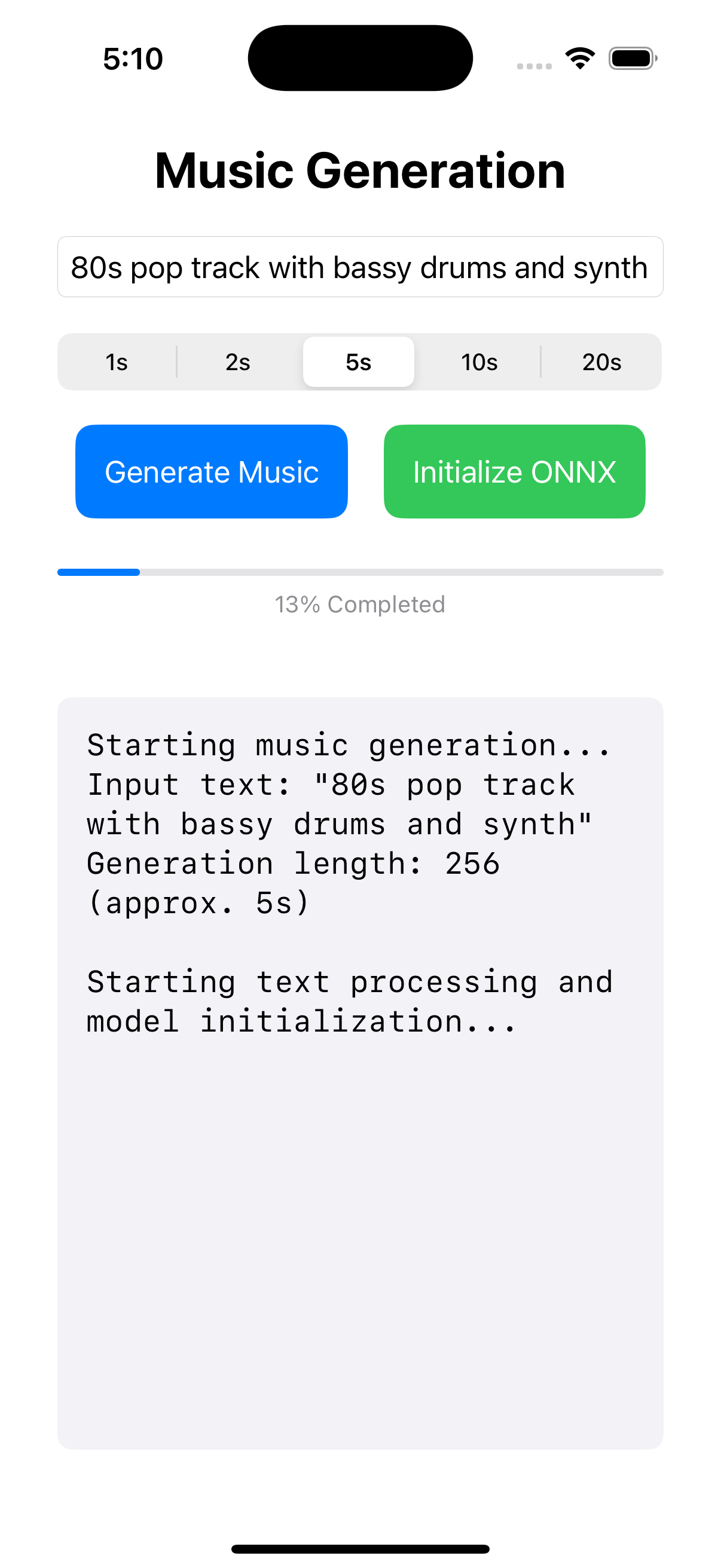} 
        \caption{Generating}
        \label{fig:sub2}
    \end{subfigure}
    \hfill
    \begin{subfigure}[b]{0.3\textwidth}
        \centering
        \includegraphics[width=\textwidth]{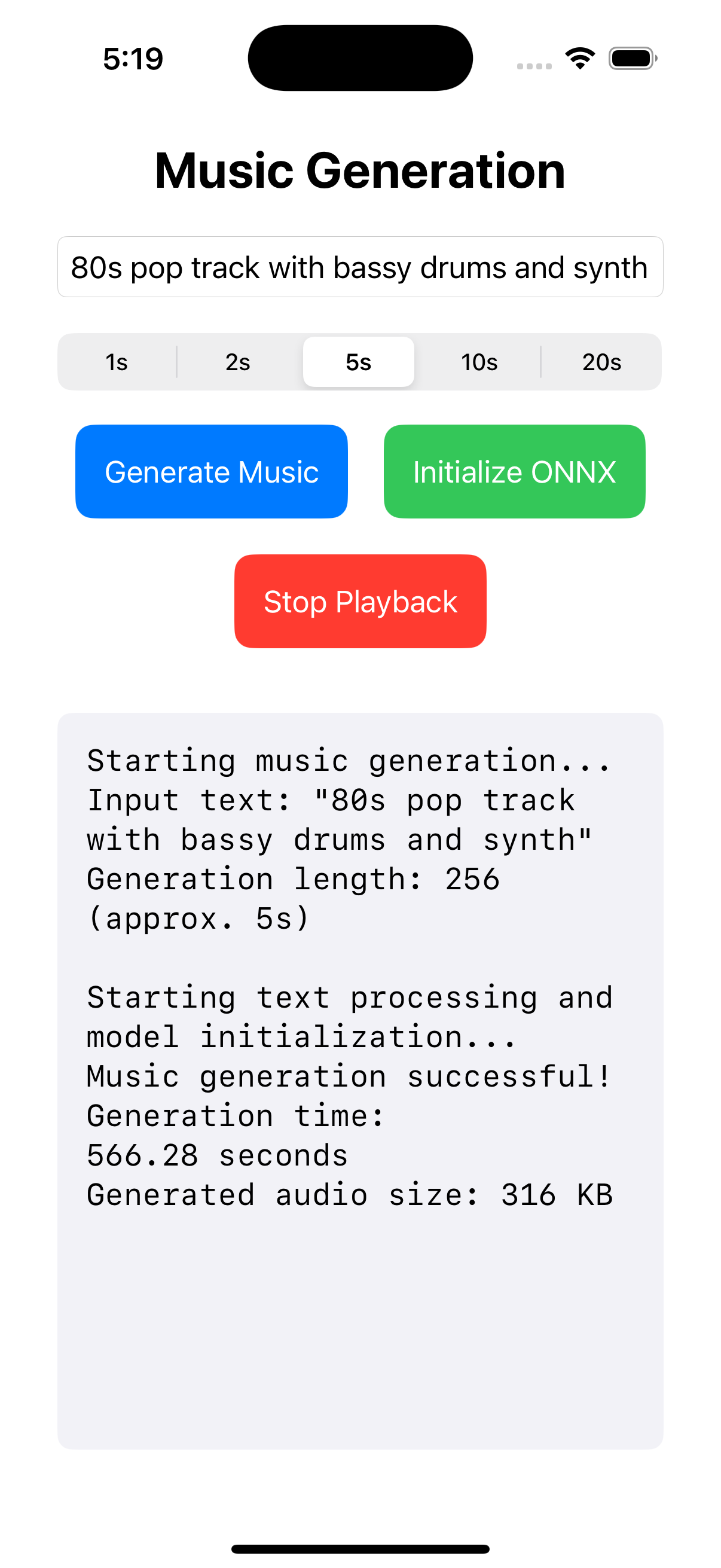} 
        \caption{Generate Complete} 
        \label{fig:sub3}
    \end{subfigure}
    \caption{iOS Music Generation App Overview} 
    \label{fig:overall}
\end{figure}

We use Xcode to develop a music generation app and deploy an ONNX model on devices by using the ONNXRuntime package. The screen is shown in Figure \ref{fig:overall}. Figure \ref{fig:sub1} (a) presents the main screen of the app. Figure \ref{fig:sub1} (b) shows the music generation process, and (c) provides the details after generating.

\end{document}

%% file: math_commands.tex
\usepackage{amsmath,amsfonts,bm}

\def\eqref#1{equation~\ref{#1}}

\def\1{\bm{1}}

\DeclareMathAlphabet{\mathsfit}{\encodingdefault}{\sfdefault}{m}{sl}
\SetMathAlphabet{\mathsfit}{bold}{\encodingdefault}{\sfdefault}{bx}{n}

